\documentclass[%
 reprint,
 amsmath,amssymb,
 aps,
 prx,
 showpacs,
 amsmath,amssymb,
 aps,
longbibliography,
superscriptaddress
]{revtex4-2}

\usepackage{graphicx}
\usepackage{dcolumn}
\usepackage{bm}
\usepackage{xcolor}
\usepackage{url}
\usepackage{breakurl}
\usepackage[breaklinks]{hyperref}

\begin{document}

\preprint{APS/123-QED}

\title{Clogging transition and anomalous transport in driven suspensions in a disordered medium}

\author{Sergi G. Leyva}
 \altaffiliation{ sergi.granados@ub.edu }
\author{Ignacio Pagonabarraga}%
 \email{ipagonabarraga@ub.edu}
\affiliation{Departament de F\'{\i}sica de la Mat\`eria Condensada, Universitat de Barcelona, Carrer de Mart\'{\i} i Franqu\'es 1, 08028 Barcelona, Spain}
\affiliation{  Universitat de Barcelona Institute of Complex Systems (UBICS), Universitat de Barcelona, 08028 Barcelona, Spain}



\date{\today}

\begin{abstract}
We study computationally the dynamics of forced, Brownian  particles through a disordered system. As the concentration of mobile particles and/or fixed obstacles increase, we characterize the different regimes of flow and address how clogging  develops. 
We show that clogging is preceded by a wide region of anomalous transport, characterized by a power law decay of intermittent bursts. 
We analyze the velocity distribution of the moving particles and   show that this abnormal flow region is characterized by a coexistence between mobile and arrested particles,  and their relative  populations change smoothly as clogging is approached. The comparison of the regimes of anomalous transport and clogging with the corresponding scenarios  of particles pushed through a single bottleneck  show qualitatively the same trends highlighting the generality of the transport regimes  leading to clogging. 
\end{abstract}

\maketitle


\section{\label{sec:level1}Introduction}
Transport in disordered media can lead to a rich  phenomenology, where particles  dynamically move freely, get trapped, and are eventually released~\cite{nuno_disordered}. 
Understanding the foundations and controlling the characteristics of clogging  and its effects is an outstanding challenge with a large number of  practical implications as diverse as  human pedestrian crowds \cite{pedestrian_review,Nicolas_2018,Yano_2022,HAGHANI2018,VONKRUCHTEN2017}, sheep herds, \cite{sheep_2015,sheep_2}, silo discharges  \cite{free_arch_2015,silo_2,zuriguel_2018}, and bottlenecks in microfluidic devices~\cite{micro_2019,sendekie_2016,laar_2016,leyva_2020}. 
In microfluidics, much effort has been taken to understand how clogging can be prevented  to avoid blocking of capillaries and develop efficient biological and medical applications in the microscale~\cite{anke_clogging,alban_2018}. Clogging is typically characterised when particles are forced to pass through a bottleneck consisting of a narrow constriction~\cite{desoir_microfluidics}. The role of the geometry, the particle shape, and the hydrodynamic coupling to  the induced flows~\cite{cruz_hidalgo_pre} has started to be analyzed systematically~\cite{schofield_clogg_flow,marin_JFM_2022}.  
Quantitative analysis of clogging in single bottlenecks 
can be successfully carried out  by measuring the difference
of the passage times between consecutive particles~\cite{clogging_zuriguel}. Its complementary cumulative distribution function (CCDF) follows a power law decay, and the tail gives the information of whether the average time
of passing particles is diverging, depending on the tail exponent, $\tau^{-\alpha}$; specifically  $\alpha<{2}$  corresponds to clogged, and $\alpha>{2}$ unclogged regimes. This exponent, hence, predicts the possibility that a bottleneck develops a clog for an indefinite period of time.
A well-known, counterintuitive observation in the passage through a constriction, and that can be quantified with this methodology is the Faster is Slower effect, in which faster entities rushing into a bottleneck results in a more persistent clogged state \cite{garcimartin_2014}. Experimental results using this approach to clogging show that the  coupling of the moving particles to the environment, e.g. through  hydrodynamics, may  affect the nature of the clogging transition~\cite{Marin_Iker_PRE}.

Clogging can also take place in a disordered system consisting of a landscape of pinned obstacles and free moving particles \cite{skyrmions_2022,reichardt_2021,reichardt_2021_2,pietro_2018,Reichhardt_2018,pietro_2018}. 
The characterisation of the filtration properties of granular media   \cite{ZAMANI200971} constitutes a relevant problem in clogging, where the goal is to minimise the flow of suspensions to filtrate a fluid, or to selectively target some specific component of the solute through a disordered medium. 
In heterogeneous environments, \citet{clogging_jamming} showed  that a completely clogged disordered landscape is characterised by a critical obstacle density, $\phi_{\text{pin}}^c$,  independent of the density of moving obstacles, $\phi_{{mov}}$, indicating that the transition to clogging is controlled by an average obstacles spacing, $l_c$. 
Furthermore, compared to jamming, clogging is characterized by  a long transient in which particles reorganize in  clogged regions of different size, leading to heterogeneous  spatial morphologies, characterized by large concentration fluctuations.

The transition of the system from  regular flow to the fully clogged regime, where there is strictly no flow, is characterized by a wide regime where flows are intermittent. 
This intermediate region is specific of the clogging transition.    
When clogging happens in local regions around an obstacle configuration, burst-like dynamics will also eventually appear and affect the flow before the whole system is clogged. Even if the average flow measured in the landscape does not vanish, locally clogged regions will coexist with free flows around other obstacles. In such intermediate states the flow is locally ill-defined, since the average time to leave a certain bottleneck may diverge. Linking the dynamics of particles travelling across a disordered landscape and local  clogging requires specific measurements that quantify whether clogs exist in a certain landscape before the average velocity vanishes and the system is fully arrested.

Understanding  clogging in a disordered heterogeneous medium will benefit from a perspective based in the adopted methodology for a single bottleneck. 
In this paper we focus on the dynamical properties of steady states of moving particles driven through an heterogeneous landscape. Our scope lays in identifying what these dynamical states consist of, and how to locally identify clogged regions when these coexist with free flow regions, bridging two complementary perspectives on the same phenomenon. 
For this purpose, we define a temporal quantity that follows the  standard methodology developed for the determination of clogged states in single bottlenecks introduced  in Ref.\cite{clogging_zuriguel}.
This temporal quantity allows us to determine whether a general landscape contains local clogs or not. 
We refer to these states with local clogs and non-vanishing average velocities in the landscape as  abnormal flow states. 
We demonstrate that in such anomalous flows, structural properties of the system such as the cluster distribution, or the velocity distribution of particles in clusters change qualitatively. By describing these dynamical  and structural properties we provide a framework to understand how local clogs rise in disordered landscape and eventually lead to total clogging of the system with increasing obstacle density. 
Similarly to fully clogged states, abnormal flow appears at a rather constant density of obstacles $\phi_{\text{pin}}^a$, which suggests the existence of an additional lengthscale that favours local clog formation. We focus on steady states of the system, where that average velocity and average cluster size have reached a steady value. This is crucial since clogging characterisation requires of long simulation runs to accurately capture the tails of power law distributions. This corresponds to relevant experimental situations, where clogs persist for arbitrarily long times compared to an initial transient state.  
We also demonstrate that power law exponents measured through single particle characteristic times are correlated to local measures of clogging as defined in the usual way
~\cite{clogging_zuriguel}.

 We structure the paper as follows: In section~\ref{sec:level2} the simulation procedure is introduced, and the magnitudes of interest are defined.
In section~\ref{sec:level3} the flow states are quantified as a function of the concentrations of  moving and obstacle particles, by calculating the complementary  cumulative distribution function of  passing times of moving particles, building on the procedure introduced to analyze clogging through a single constriction~\cite{clogging_zuriguel}.
This methodology allows to introduce a general notion of abnormal flow, where localised flow of particles coexist with persistently clogged regions where the flow is not well defined. 
This new flowing regime allows to build a state diagram that distinguishes between normal flow, abnormal flow, and clogged states, where the average velocity is zero. In section~\ref{sec:level4} we compare the developed methodology with a local measure of clogging and establish a clear correlation between both approaches. Thus, we confirm the intuition that the abnormal region is a consequence of locally constricted regions, and identify the same trends and clogging exponents for both methodologies. 
In section~\ref{sec:level5}, the dynamic and structural features of normal and anomalous flows are compared. We characterise the distribution of clusters size and the probability distribution functions of the velocity of particles belonging to clusters interacting with obstacles, which show that such quantities depend strongly on the system density.   We finish with the main conclusions and implications of the obtained results in section~\ref{sec:level6}.

\section{\label{sec:level2}Simulation methodology}

We carry out Brownian dynamics simulations of a 2D system of area $L^2$ with periodic boundary conditions, composed by a  total number of $N=N_{mov}+N_{pin}$ disks of radius $\sigma$. $N_{mov}$ disks move under the action of forces, while $N_{pin}$ remain pinned at their  initial positions.  
Both moving and pinned particles interact sterically with a force that derives from a Yukawa potential \cite{Lowen_1993}

\begin{equation}
\label{eq:yuk}
{\bf{F}}^{int}({\bf{r}}_{ij})=\frac{U_0}{\lambda}\frac{\sigma}{r_{ij}}\left(\frac{\sigma}{r_{ij}}+\frac{\sigma}{\lambda}e^{\frac{-r_{ij}}{\lambda}}-B\right)\hat{r}_{ij},
\end{equation}

\noindent where ${\bf{r}}_{ij}$=${\bf{r}}_i-{\bf{r}}_j$, $r_{ij}=|{\bf{r}}_i-{\bf{r}}_j|$ and $\hat{r}_{ij}={\bf{r}}_{ij}/r_{ij}$, where $i$ and $j$  refer to both moving and pinned particles. The parameter $\lambda$ characterises the decay range of the steric interaction, while $U_0$ is the interaction strength.

Moving particle $i$ evolves according to an overdamped dynamics

\begin{equation}
\label{eq:brownian}
\frac{1}{\mu}\frac{d{\bf{r}_i}}{dt}=\sum_{j\neq i }^N{\bf{F}}^{int}({\bf{r}}_{ij})+{\bf{F}}^{ext}+{\bf{F^T}}({\bf{r}}_i).
\end{equation} 

\noindent where, $\mu$ is the  disk mobility and relates the short time diffusion coefficient and the temperature through the Boltzmann constant, $D_0=\mu{k_BT}$.   

 The driving force has a constant value and without lose of generality, is chosen to act on the x-direction so that ${\bf{F}}^{ext}=F_{D}\:\hat{{\bf{x}}}$, where $\hat{{\bf{x}}}$. The last term in Eq. \ref{eq:brownian}  accounts for the thermal bath, and its integration over a time step describes a Gaussian random displacement with second moment $\Delta {\bf{r}}=\mu {\bf{F}}^T_i\Delta{t}$ such that $<(\Delta{r^T})^2>=2D_0\Delta{t}$ and zero mean.

The dynamics can be expressed in dimensionless form scaling distance and time by appropriate reference quantities. We consider the particle radius, $\sigma$, as the characteristic distance, and  the characteristic time as the time required for a particle dragged by the driving force to move its own radius, $\tau_D=\sigma/(\mu{F_D})$. Accordingly, Eq.~(\ref{eq:brownian}) reads

%
\begin{figure}[b!]
\begin{flushleft}
 {(a)} 
\end{flushleft}
\begin{center}
\includegraphics[width=0.85\columnwidth]{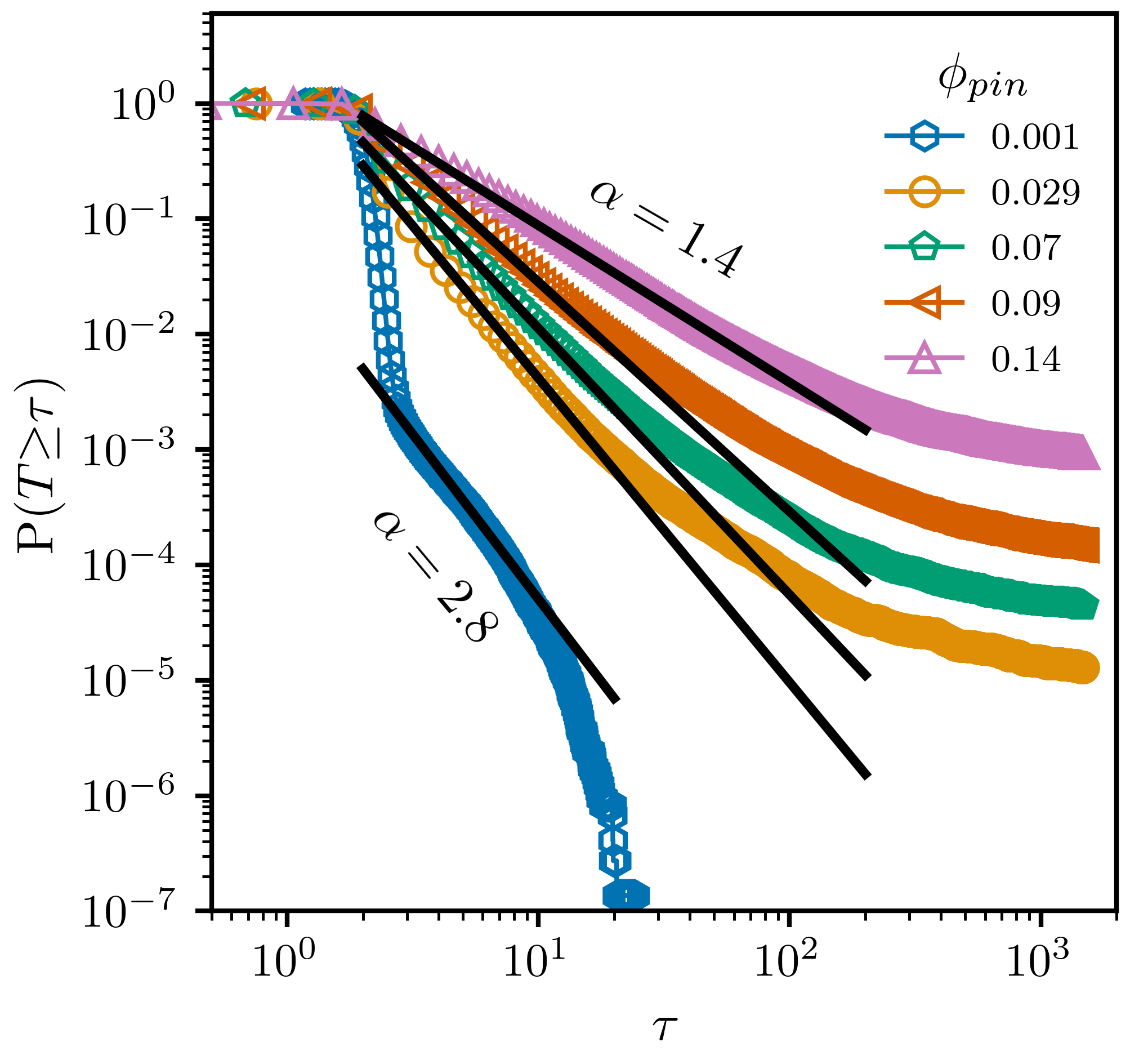}
\begin{flushleft}
 {(b)} 
 \end{flushleft}
\includegraphics[width=0.85\columnwidth]{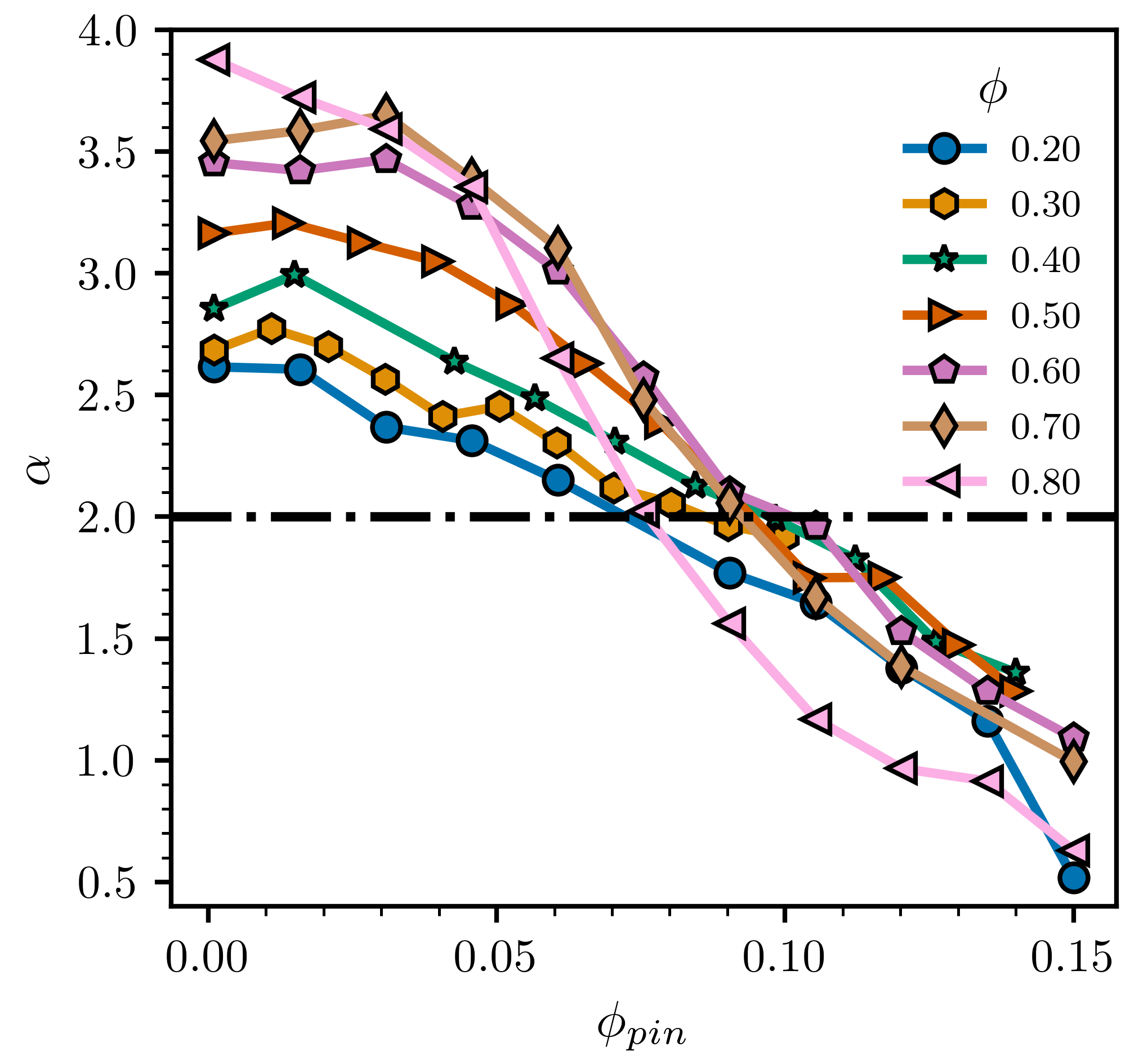}
\end{center}
	\caption{ (a) Computed complementary cumulative distribution function  (CCDF) for $\phi=0.40$. Black lines corresponds to the  power law fit, and the  calculated exponent, $\alpha$ is shown for the two extreme cases considered. The decay tipically starts fot $T>2$, since it is the minimum moving time of a free particle, according to our definition of an event. Depending on the fraction of obstacles, one can observe a fast decay region produced by a majority of moving particles, or a power law region, where particles often interact with obstacles, which can lead to clogging events.   (b) Characterisation of $\alpha$ as a function of the immobile particle packing  fraction, $\phi_{pin}$, for different total packing fractions, $\phi$.   
	\label{figure1}}
\end{figure}
\begin{equation}
\label{eq:brownian_non_dim}
\frac{d{\bf{\bar{r}}_i}}{d\bar{t}}=\frac{U_0}{\lambda{F_D}}\bar{\psi}({{\bf{\bar{r}}}_i}/{\sigma},{\lambda}/{\sigma})\hat{{\bf{r}}}_{ij}+{\bf{\hat{x}}}+\sqrt{\frac{2 \tau_D}{Pe}}{\bm{\bar{\xi}}},
\end{equation}
\noindent where bar indicates that the magnitudes have no dimensions.
The term $\bar{\psi}({{\bf{\bar{r}}}_i}/{\sigma},{\lambda}/{\sigma})$ is the Yukawa force in Eq.~(\ref{eq:yuk}) divided by $U_0/\lambda$. The term ${\bm{\bar{\xi}}}$ describes a Gaussian stochastic function with  $<{\bm{\bar{\xi}}}>=0$ and $<({\bm{\bar{\xi}}})^2>=1$. The P\'eclet number, $Pe=v_0\sigma/D_0$,  quantifies the ratio between the velocity and the thermal contribution to the  particle motion.

We are interested in the regime where driving and inter-particle  forces dominate over  thermal fluctuations. Accordingly, we consider $U_0/\lambda{F_D}=300$, $Pe=100$ and $\lambda/\sigma=1$. The time step, $\Delta{t}$, is chosen small enough to avoid particle overlapping, $\Delta{t}/\tau_D=1\cdot10^{-3}$.  Both moving and pinned particles are initialized following a growing algorithm in which particles and obstacles are placed randomly in space  and then evolved in time to grow to its size to reach the desired area fraction~\cite{clogging_jamming}.

The number of moving particles in the simulation is constant and large enough to provide reliable statistics, $N_{mov}=10000$. Simulations are performed fixing the   total packing fraction, $\phi=(N_{\text{mov}}+N_{\text{pin}})\pi\sigma^2/L^2$, and  varying the pinned packing fraction of particles ,$\phi_{\text{pin}}=N_\text{{pin}}\phi\sigma^2/L^2$. 
Hence,  the state of the flow will be characterized as a function of $\phi_{\text{pin}}$ and $\phi$. 

The simulation is first run until the average velocity of moving particles becomes constant $\left<v_x\right>=cte$ . After this initialisation,   the tendency of  particles to flow is captured by means of a characteristic time quantity , $\tau$, that we explain  below. In order to compare with standard clogging measurements, we will also measure the average particle velocity, and identify a state as clogged if it  exhibits a zero average velocity in the direction of the driving force, thus $<v_x>=\sum_{i} v_{i,x}\simeq{0}$. 
Computationally,  we never observe  $<v_x>\simeq{10^{-5}}$ due to the thermal fluctuations. Thus, we take this threshold to identify a fully clogged state.  

\section{\label{sec:level3}Flowing states }

We characterize the state of flow of the system computing the CCDF of disk displacement times. This function is constructed by   quantifying the time, $\tau$, it takes a disk to displace its own diameter, $d=2\sigma$, in the direction of the driving force. We identify such intervals through dynamical measurements (DM),  where we  identify  all events  in which any given disk has moved a distance $d$ through the numerical integration of  Eq.~(\ref{eq:brownian_non_dim})~\footnote{Specifically, we follow the displacement of all particles and, when  the displacement reaches $d$, we identfiy the associated time, $\tau$, and reset the corresponding counter of the particle to 0}.
These events allow to determine the  dynamic regimes of the  moving disks. For example, free flowing particles have passing times close to ${\tau}\sim2\tau_D$, while particles that interact with obstacles will exhibit  larger $\tau$. 
The flow regimes of the forced suspension are  then analyzed using the CCDF, $P(T > \tau)$ that quantifies the fraction of  all events that take a time $T$ larger than a prescribed value $\tau$. 
Later we will also characterise such events with static measurements (SM),  a procedure that is equivalent to the standard local characterisation of clogging through bottlenecks.

Fig.~\ref{figure1}a displays the CCDFs for  a given overall area fraction, $\phi=0.4$ as a function of the fraction of pinned disks, $\phi_{pin}$. One can identify three different dynamical regimes.
A  first region of fast decay near $T=2$ is observed for small fraction of obstacles, which corresponds to particles which do not  interact  strongly with  the obstacles and are essentially driven  by the applied force at constant velocity. At larger times, a second region generally appears, characterized by larger displacement times, which is due to the interaction of the driven particles with fixed obstacles.  This region can be characterized by a power law decay: Moving particles interact with obstacles, become trapped and may be able to move eventually. These interactions with constrictions and other free particles can give rise to clogging events that persist in time in certain bottlenecks of the system.
For increasing obstacle fraction, the decay of the CCDF can start with this second region, as observed in Fig.~\ref{figure1}a.
Finally, a third region appears at largest times, produced by obstacles, in which the power law behaviour is lost.  For such large times, the  deviation is produced by particles that remain blocked most of the simulation run, typically due to a geometric confinement that hinders the flow, with no unclogging possibility. 
The saturation of the CCDF observed for increasing $\phi_{\text{pin}}$ is due to such blockage of free particles.

As shown in Fig.~\ref{figure1}a, the second region can be adjusted by a power law, and the corresponding clogging exponent $\alpha$ can be systematically obtained following the procedure stated in Ref.~\cite{newman_power}, as a function of  $\phi_{\text{pin}}$.
If $\alpha{\leq}2$, the average passing of particles diverges, which means that in some regions of the landscape a clog can exist for an indefinite period of time, and will thus result in a local accumulation of particles. 
In such cases, the system may not be fully clogged and its average velocity may not be zero, but clogs coexists with flowing states of particles.  
Thus, in general, in this regime the average flow of particles can be well defined only locally in some regions of the landscape; accordingly, we refer to this flowing regime as abnormal flow.

By calculating the $\phi_{pin}$ at which the power law diverges,  $\alpha(\phi_{pin}^a)=2$, we can characterize the fraction of obstacles at which such abnormal flows are developed.  
Fig.~\ref{figure1}b), displays the value of $\alpha$ as a function of $\phi_{\text{pin}}$ for different $\phi$, and we find an important feature of clogging of colloidal suspensions in disordered media: 
The fraction of obstacles where normal flow becomes abnormal remains roughly constant $\phi_{\text{pin}}^a\simeq{0.09}$,  with a weak dependence on the overall  area fraction $\phi$.

\begin{figure}[t!]
\includegraphics[width=3 in]{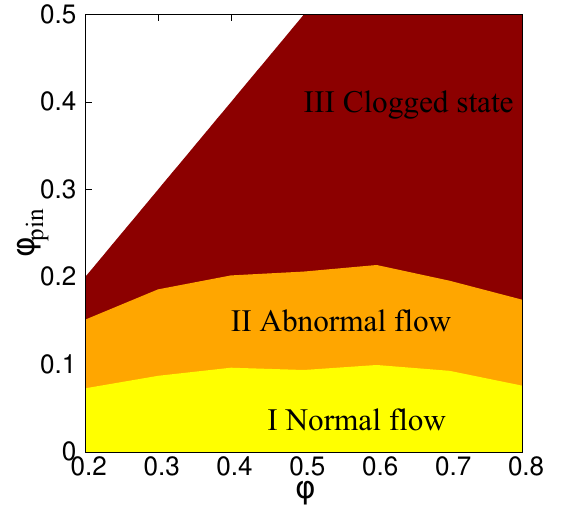}
\caption{ State diagram, which identifies the three regimes of  collective particle displacement of normal flow, abnormal flow,  and clogging. The maximum width of the normal region is observed for intermediate densities, while it decreases for small $\phi$, where isolated particles get trapped easily in constrictions, or for large $\phi$, where when we approach the jamming transition.}	\label{figure2}
\end{figure}

The different dynamical regimes that control the transition from normal flow to  complete clogging  for the driven disks in a system composed  by a random distribution of non-overlapping obstacles can be summarized in the state diagram of Fig.~\ref{figure2} that identifies the region of normal flow ($\alpha>2,{\langle}v_x{\rangle}>0)$), abnormal flow ($\alpha<2,{\langle}v_x{\rangle}>0)$)), and clogging ($\alpha<2,{\langle}v_x{\rangle}=0$), as a function of $\phi$ and $\phi_{\text{pin}}$. This diagram is similar to that shown by \citet{clogging_jamming}: We see that flow vanishes at a constant critical obstacle density $\phi_{\text{pin}}^c$. This constant $\phi_{\text{pin}}^c$ indicates the existence of a characteristic distance $l_c$ between obstacles that impedes particle flows. 
Introducing in the diagram the notion of abnormal flow, we observe an additional anomalous region where the average velocity is not strictly zero, yet we observe that the distribution of times required for a particle to move its own diameter $\tau$ is diverging. In these states, clogs can locally develop for an indefinite period of time, dramatically altering the flowing properties of the moving particles in the landscape. 
Furthermore, we observe that similar to $\phi_{\text{pin}}^c$, the critical abnormal flow density $\phi_{\text{pin}}^a$ also depends weakly on the obstacle density, suggesting an additional characteristic distance between obstacles $l_a$ that sets the appearance of local clogs in the landscape. Different steric potentials will affect these  characteristic sizes, $l_c$ and $l_a$, that sets the diagram width, but will not affect the observed phenomenology. It is true that significant changes in the character of the potential, e.g. its  range and attractive nature, can  affect the stability of the  clusters and clogs significantly. Nonetheless, these aspects  complement the main message of this piece of work and may be the subject of subsequent research.

\begin{figure*}[t!]
\includegraphics[width=\textwidth,height=11.3333cm]{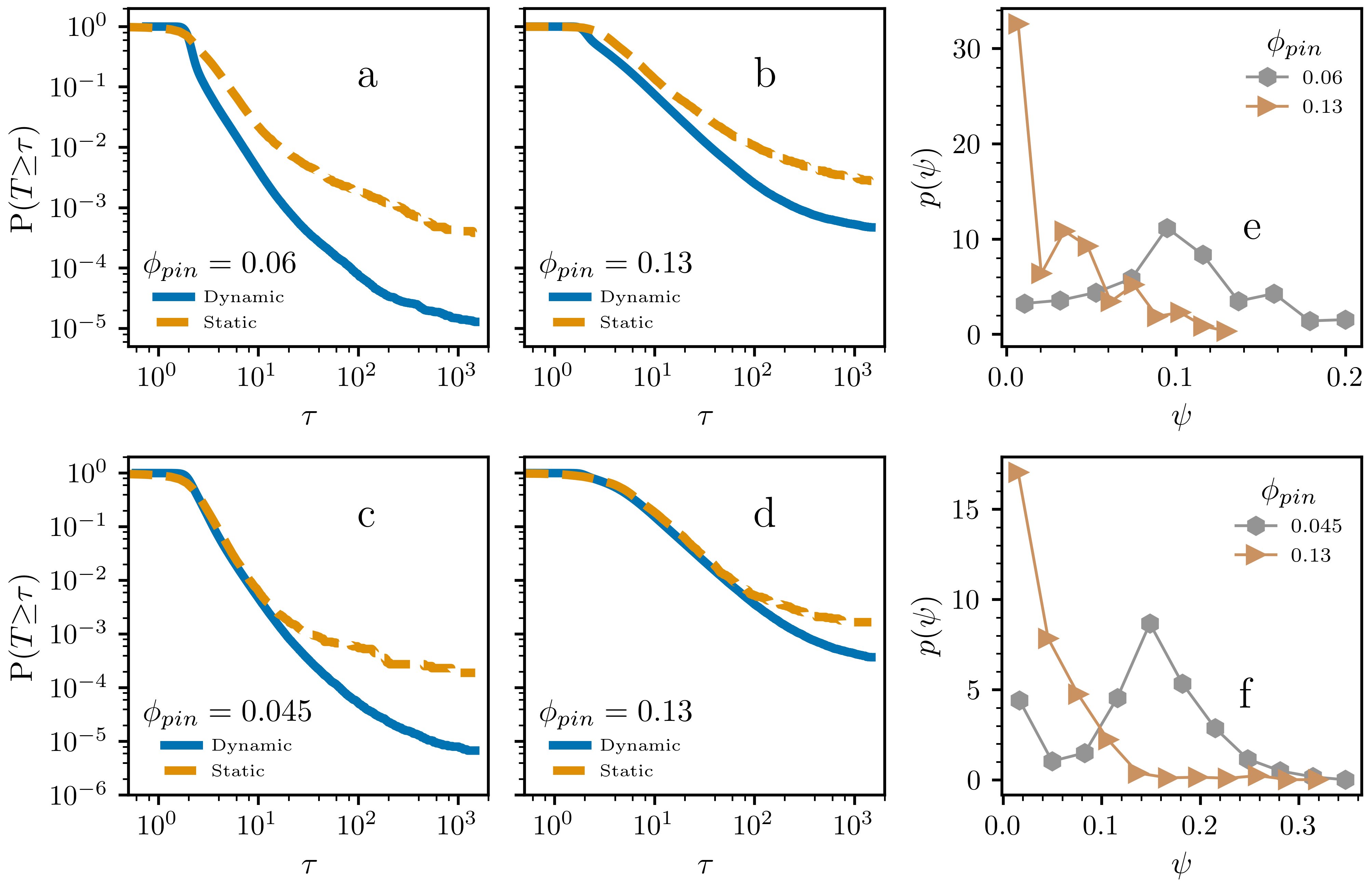}
	\caption{\label{figure3} SM and DM comparison for $\phi=0.4$ (a,b,e) and $\phi=0.6$ (c,d,f), in the normal region (a,c) and abnormal region (b,d). Our results show how SM and DM are not independent of each other and correlate in the characteristic algebraic decaying exponent. (e) and (f) show how both densities exhibit the same tendency: In the normal region, the flow pdfs peaks at a maximum value at the center, while in the abnormal region the flow pdfs has a maximum value at 0 and decreases with increasing flow rate. }
\end{figure*}

The average height of the of the normal flow region is around $\phi_{\text{pin}}=0.09$, which is a 
relatively small area fraction. 
The normal and abnormal regions are comparable in width, showing that anomalous flow is not a marginal feature that takes place right before reaching a completely clogged state. Aditionally, in the diagram, we find evidence of cooperation, as for increasing density $\phi$, the normal region becomes thicker: For a constant $\phi_{pin}$ we can eliminate local clogs by means of increasing the fraction of moving particles. 

At high densities, $\phi\ge{0.65}$, we expect that  increasing $\phi_\text{pin}$ the system exhibits jamming~\cite{clogging_jamming}. Some features indeed point towards the existence of the jamming transition in these regions: Both $\phi_{\text{pin}}^a$ and $\phi_{\text{pin}}^c$ slightly decrease for increasing $\phi$. 
Even before the jamming transition, a region of abnormal flow develops before the average velocity decreases to zero $\left<v_x\right>=0$. Hence, the anomalous flow regimes are a general, strong feature of disordered landscapes, that can smoothly lead to fully clogged states as the fraction of obstacles increase. 

In the next section we will establish the connection between clogging measured as  previously described, and clogging measured in local regions of the landscape, which play a similar role of a  bottleneck. 

\section{\label{sec:level4}Local flow properties}
To provide further  insight on the implications of the local spatial organization of abnormal flowing events,  we analyse the flow of particles and compare the  clogging measurements as typically measured locally through bottlenecks \cite{clogging_zuriguel}. For this purpose, we divide the simulation box in the $y$ direction in sections of a characteristic width $l_s$. 
We choose  $l_s = 2.5\sigma$ comparable to $l_c$, which is of the order of magnitude of particle dimension.  We measure the  time interval it takes   two consecutive particles to cross the region defined by $l_s$~\cite{clogging_zuriguel}. We shall refer here to this procedure as static measurement (SM), as opposed to the previously  DM protocol.

\begin{figure}[h!]
\includegraphics[width=\columnwidth,height=8.5cm]{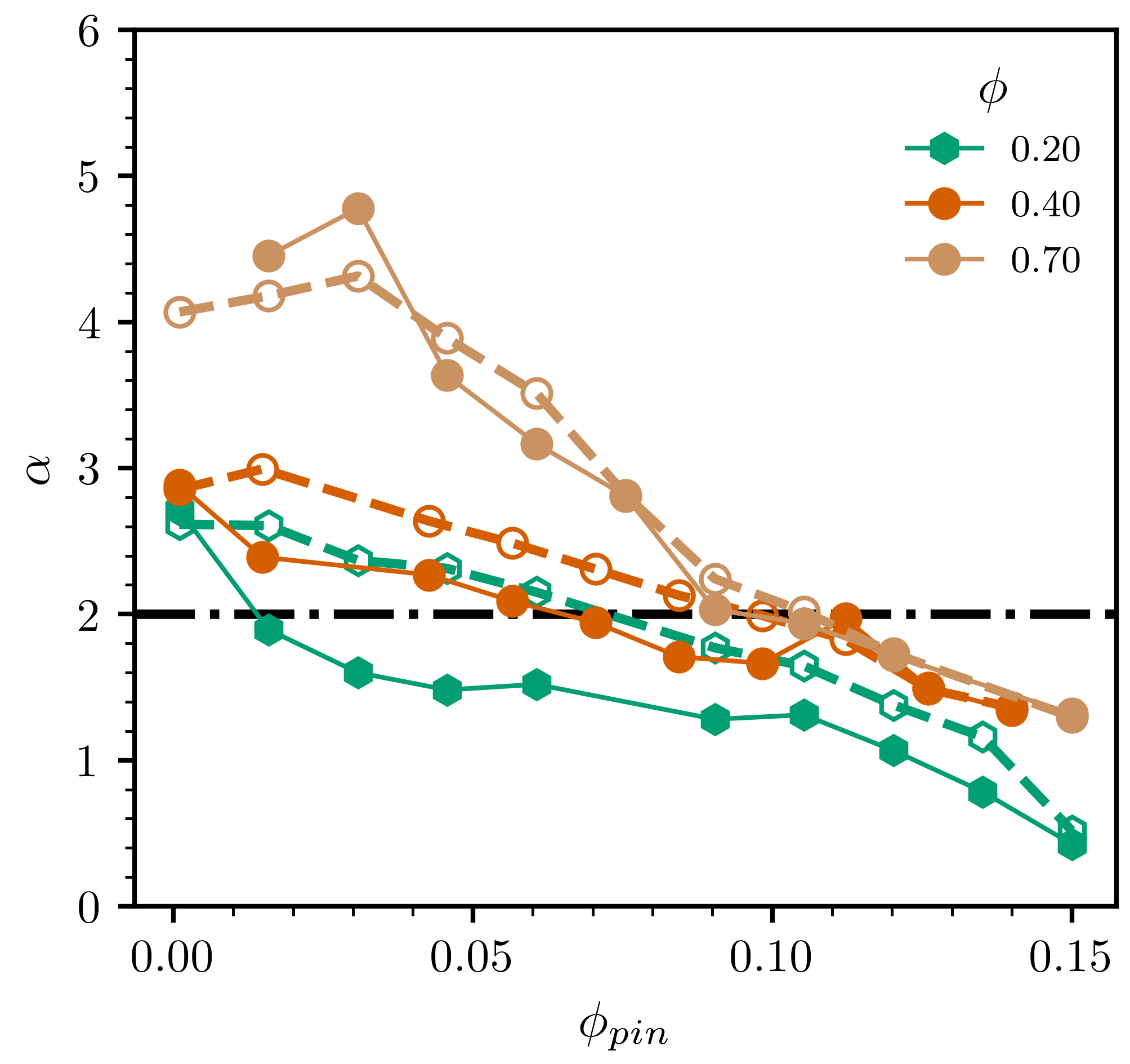}
	\caption{\label{figure4} Static measurements (filled markers with continuous lines) vs dynamic measurements (dashed lines and empty markers), for different total area fractions. Both methods exhibit qualitatively similar exponents. The SM method tends to overestimate $\alpha$ compared to DM. This tendency weakens as $\phi$ increases, and all the moving particles connect forming a large continuous cluster.   }
\end{figure}

Fig.~\ref{figure3}.a-d  displays the CCDF obtained using the SM and DM protocols. The curves show similar trends, and indicates that SM sistematically overestimates the  events that require larger times, hence underestimating the value of $\alpha$. In general, this deviation  decreases with  increasing $\phi_{pin}$ and $\phi$, as can be appreciated in Fig.~\ref{figure4}. The underestimation of $\alpha$ using SM is due to its sensitivity  to flow disturbances  due to large passing times produced by density fluctuations. For $\phi=0.6$ we observe that both methods  give quantitatively similar exponents. Thus, we find that  SM and DM provide complementary  methods to analyze the emergence of abnormal flow in suspensions of forced particles in a disordered system. 
In other words, locally measuring the flow along the disordered obstacle at a certain definite locations, is akin to following the flow of individual particles. 
However, the DM method, when characterising the dynamic properties of a certain disordered medium consisting of an arbitrary array of constrictions, provides a more robust characterisation because it is less sensitive to obstacle density fluctuations. 

The robustness of the measured exponent $\alpha$ suggests that the   state diagram, and the  presence of an abnormal flow regime initially identified for  the flow through  isolated bottlenecks is  a generic feature of the clogging transition. The comparison between DM and SM  provides complementary strategies to analyze the transition when clogging does not take place through a unique obstacle.

We can also quantify the  local particle flow using  SM. To this end, we count the  number of  particles crossing  a prescribed segment of length $l_s$ perpendicular to the direction of the driving force  during a prescribed time interval  $\Delta{t}$. We choose $\Delta{t}=20{d}\tau_D$, as a compromise to analyse the flow during a relevant amount of time minimising the impact of dispersion  due to individual particle motion. The flow in each cross section, defined by $l_c$, is then calculated as $\psi=n_{mov}/l_c\Delta{t}$, where $n_{mov}$ is the number of moving particles across the line defined by $l_{c}$ in a time $\Delta{t}$. 
Fig.~\ref{figure3}e,f shows that in the normal regime, the flow peaks around a certain value that depends on $\phi_{\text{mov}}$. In the abnormal regime, the flow distribution decreases monotonously and has its maximum  at $\phi=0$, providing a complementary perspective on the properties of the abnormal flow as opposed to normal flow.

In this section we have explored the clog dynamics in local regions of the landscape, in a similar way as it is tipically characterised in single bottlenecks \cite{clogging_zuriguel}. Our analysis demonstrates that the anomalous dynamics observed  is directly correlated to clog development in localised regions of the system. 
The picture that emerges  is that of a disordered system with heterogeneous dynamics, where bottlenecks with diverging distributions of characteristic passing times, $\tau$, coexist with free flow paths \cite{EPAPS}. These bottlenecks  tend to accumulate particles, forming large dense clogs, and flow will tend to be localised around these bottleneck zones or regions where fluctuations of obstacles density have allowed a constant flow . 
To complete this picture, we now turn to study the distribution of particle velocities in clusters, and cluster distributions, in order to map this description to a geometrical and dynamical picture of the properties of moving particles in these anomalous flow regimes.  

\section{\label{sec:level5} Normal and abnormal dynamics }

To gain insight on the particle dynamics that gives rise to abnormal flow and the related emergent properties, we analyze  the disk cluster distribution  and the relation to the velocity distribution  at the steady state.  We use a distance criterion, and consider that  all  particles  with a separation  smaller than $\sigma+\delta$ belong to the same cluster \footnote{We choose $\delta=0.1\sigma$, to minimize the impact  of  thermal displacements in the cluster characterization.}.  
Fig.~\ref{figure3}a shows the cluster probability distribution function (pdf) in the different flowing regimes. 
The decay  of the pdfs is generically compatible with an  algebraic decay. For small densities, {\sl e.g.} $\phi=0.2$, moving from the normal (dashed line, pink triangles) to the abnormal  (continuous line, pink pentagons) flow regime results in a slower decay of the pdf, with an effective exponent of the algebraic tail  that increases from $\xi<-2$ to $\xi>-2$. This implies that for small clogging densities, in the anomalous regime, the average number of clusters diverges and there is no characteristic cluster size. Instead, for arbitrarily large systems, and thus increasing $N$ with the same $\phi_{\text{pin}}$ and $\phi_{\text{mov}}$, we will find arbitrarily large clusters in the anomalous regime.

\begin{figure}[t!]
\begin{flushleft}
 {(a)} 
\end{flushleft}
\begin{center}
\includegraphics[width=\columnwidth]{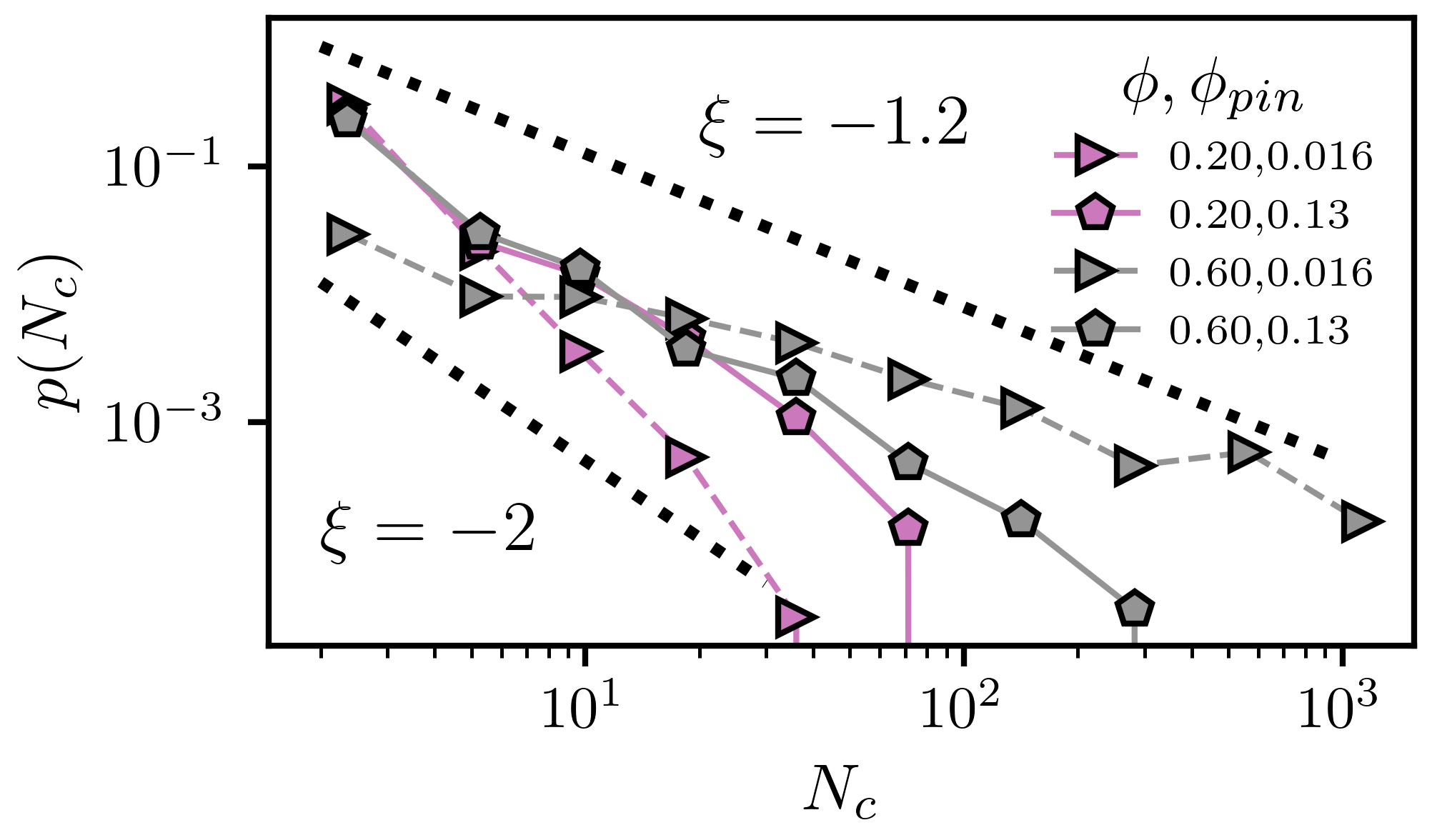}
\begin{flushleft}
 {(b)} 
 \end{flushleft}
\includegraphics[width=\columnwidth]{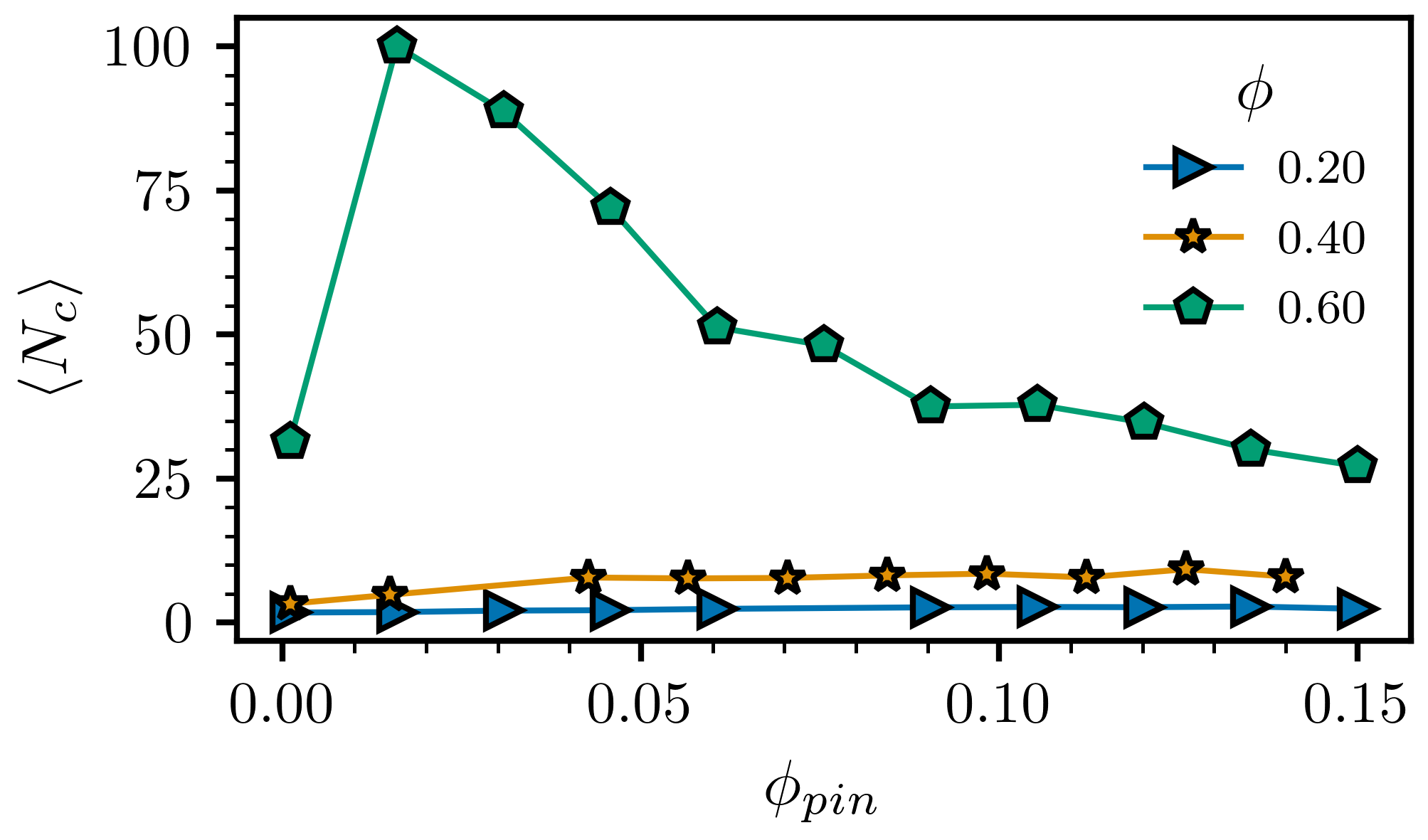}
\end{center}
	\caption{ (a) Probability distribution function of number of particles in clusters, in the normal region (dashed lines, triangles) and abnormal region (continuous lines, pentagons), for two different densities, $0.2$ (pink) and $0.6$ (silver). For $\phi=0.2$, entering the abnormal region implies an increase of algebraic exponent $\xi$, while for $\phi=0.6$ it implies a decrease of $\xi$. This highlights qualitatively  different flowing properties for small and large concentrations.     (b) Cluster size as a function of $\phi_{\text{pin}}$ for different $\phi$. For $\phi<0.5$, the average number of clusters increases gradually with increasing $\phi_{\text{pin}}$, while for $\phi>0.5$ it decreases, showing how in this case interrupting the flow translates into smaller clusters.}
	
 \label{figure5}
\end{figure}

\begin{figure*}[t!]
\includegraphics[width=\textwidth,height=11.3cm]{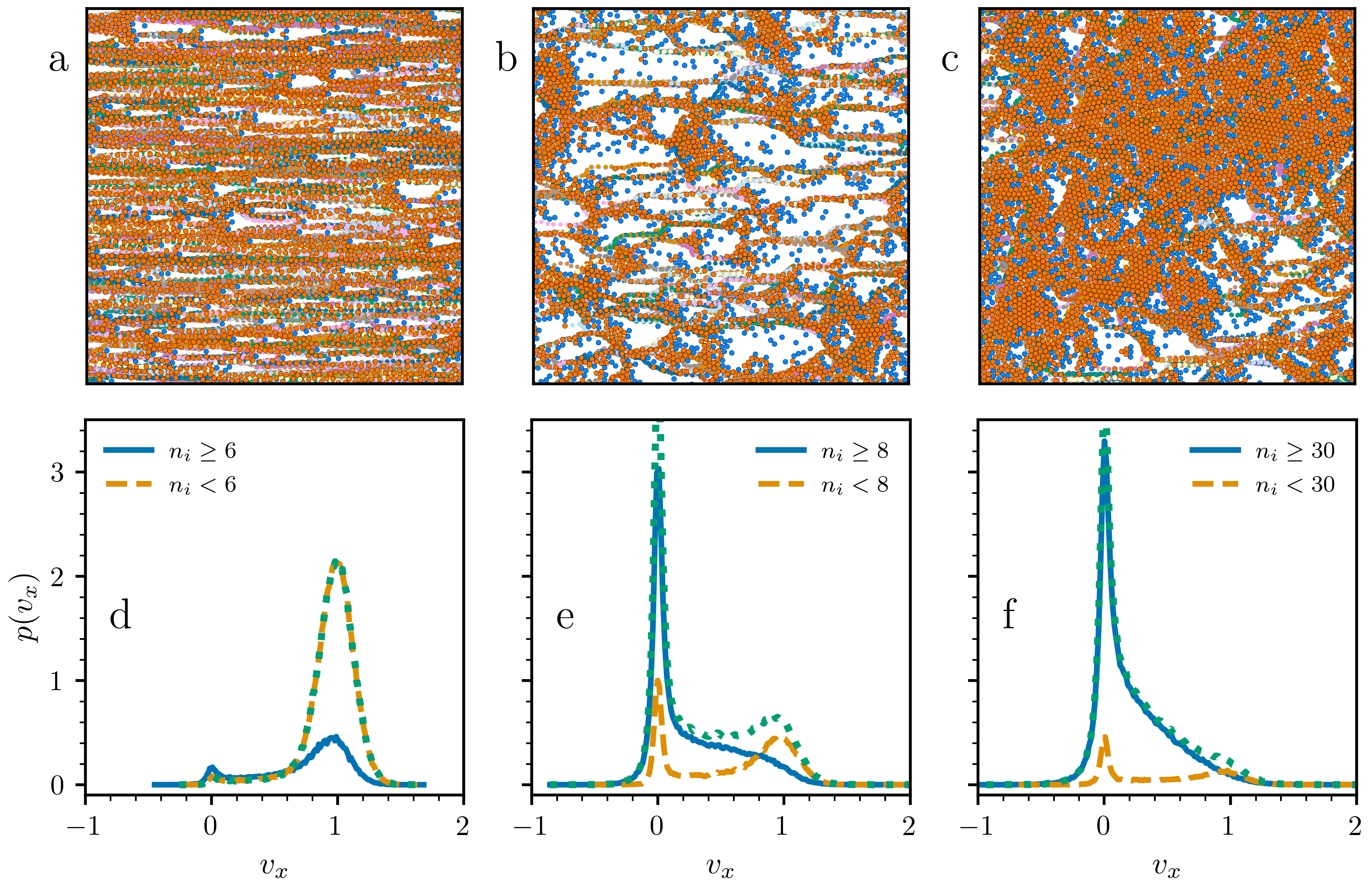}
\caption{\label{figure6} (a,d) Shows a state of normal flow with $\phi=0.4$, $\phi_{\text{pin}}=0.03$ (b,e) Shows a state of abnormal flow with $\phi=0.4$, $\phi_{pin}=0.13$ 
(c,f) Shows a state of abnormal flow with $\phi=0.6$, $\phi_{pin}=0.13$, respectively. Figs (a,b,c) show snapshots of the simulations, different coloured lines correspond to different particle trajectories, showing where the flow tipically takes place in the landscape. Figures (d, e, f) show the  velocity distribution for clusters with $n_i$ particles smaller than the average size ${\langle}N_c{\rangle}$ (dashed line),  clusters larger than ${\langle}N_c{\rangle}$ (continuous line), and the probability distribution of all the particles (dotted line) . The velocity is calculated for all particles as $v_x=d\langle{\bar{r}_x}\rangle/d\bar{t}$. For $\phi<0.5$ the abnormal region exhibits a bimodal distribution, where big clusters have have the most particles at $v_x=0$, and small ones the largest velocities peaking at $v_x=1$. For $\phi>0.5$, the doubled peaked distribution disappears, and the total pdfs almost coincide with those belonging to large clusters.}   
\end{figure*}

At higher concentrations, {\sl e.g.} $\phi=0.6$, in the normal flow regime  (dashed line, silver triangles),  in both normal and abnormal regions $\xi>-2$,
implying that the mean cluster size is always diverging. The difference remains in the fact that now, arbitrarily large clusters will appear too in the normal flow regime, but these clusters do not induce clogging. This change of trend translates into qualitative differences in the morphology and flowing characteristics of the system depending on the total fraction of particles. 
To further understand how the area fraction affects the distribution of particles in the system, in Fig.~\ref{figure3}b we show the  average number of particles in clusters ${\langle}N_c{\rangle}$ for different $\phi$, as a function of $\phi_{\text{pin}}$. Indeed, it shows a strong qualitative dependence of ${\langle}N_c{\rangle}$ for the different curves, depending on $\phi$. 
At small $\phi$, flowing particles remain in small clusters. Keeping $\phi$ constant, as   $\phi_{\text{pin}}$ increases, particles increase their probability to  accumulate in small groups near obstacles, which explains the increase in ${\langle}N_c{\rangle}$. 
For larger $\phi$, already at small $\phi_{\text{pin}}$ particles display a strong probability to accumulate  near obstacles while still being able to flow. 
By increasing $\phi_{\text{pin}}$  starting in the normal regime, obstacles initially divide the flow in disconnected regions of normal flow, sharply decreasing the average cluster size. Eventually, when $\phi_{pin}>\phi_{pin}^a$ flow is interrupted and results in clogs, as depicted in Fig.~\ref{figure6}a-b,changing the decreasing trend, since now particles are not only divided in disconnected regions but  also accumulate in clogs, as shown in Fig.~\ref{figure5}b. Hence, in this case local clogs appear, and $\left<N_c\right>$ decreases more smoothly, since such local clogs arise in spatially uncorrelated regions of the system disconnecting flow regions, but still favouring accumulation of particles in bottlenecks.
Even if the dependence of $\left<N_c\right>$  with $\phi_{pin}$ differs qualitatively for large and small $\phi$, the resulting states are the same: Particles separate in regions of high density near bottlenecks and regions of small density between bottlenecks.

To quantify the impact of the dynamic properties of the clusters on these different scenarios we  compute the velocity pdfs
of particles belonging to clusters larger and smaller than the average cluster size, ${\langle}N_c{\rangle}$, and for two different densities.
Fig.~\ref{figure6} displays a series of snapshots of the clustering of disks in the normal, Fig.~\ref{figure6}.a, and abnormal, Fig.~\ref{figure6}.b and Fig.~\ref{figure6}.c, regimes. The plots  show   that abnormal flow correlates with the development of large clusters seeded around  regions with a local enhancement in the concentration of obstacles. As the overall  packing fraction increases, Fig.~\ref{figure6}.e, the clusters grow towards a jammed state.

Fig.~\ref{figure6}.d-f displays the velocity pdfs for particles interacting with obstacles and belonging to  small and large clusters. For normal flow, Fig.~\ref{figure6}.d, most particles displace at  the velocity corresponding to free flow, $v_f=d/2\simeq{1}$, driven by the external force. Only a small fraction of the particles are trapped by an obstacle, displaying a velocity close to zero. This fraction is slightly larger for the small fraction of  disks which belong to large clusters. 

Entering the abnormal flow regime, the velocity pdfs for particles in small and large clusters  show some qualitative differences.
In the abnormal flow regime, far from the clogging transition, the velocity distribution of moving particles shows a characteristic two-peaked bimodal distribution, as observed in Fig.~\ref{figure6}e. Local clogs coexist with normal flows, as appreciated in Fig.~~\ref{figure6}.b. 
Particles belonging to small clusters exhibit clearly this two-peaked bimodal distribution, with a finite fraction of particles displacing in reaction to the applied force, $v_f$, corresponding to localised particle free flow. Particles  released from clogged states contribute to this peak, as they form trails  moving freely until reaching the next clogged region. The other peak correspond essentially to arrested particles, with a velocity close to zero.  Small clusters of particles accumulating at clog regions in specific bottlenecks of the system , temporarily or spatially isolated from flowing regions contribute to this peak . All these events can be observed in the snapshots of Fig.~\ref{figure6}.b.
In large clusters, the largest peak appears at $v_x\sim{0}$, produced by large regions where clogs persist in time, giving rise to intermitent flows, but also in coexistence with paths where particles can flow.   
Such mixed state highlights the key ingredient of abnormal flows in disordered mediums: Intermittent flows and temporary blockages arise locally throughout the disordered system as particles are dynamically trapped and released from local constrictions.

As shown in Fig.~\ref{figure6}.f, at higher $\phi$, as we approach the clogging
transition, a smaller fraction of disks  are contained within small clusters and the bimodal velocity distribution is barely visible.
Increasing $\phi$ decreases the regions of locally small density, as can be seen in Fig.~\ref{figure6}b,  favouring that all particles belong to few large clusters that dominate the system. For such large densities, 
instead of having a large number of small clusters distributed in uncorrelated bottlenecks, now we find a small number of big clusters, where the velocities inside the same cluster are correlated.
The resulting velocity distribution corresponds to the attenuation of the bimodal two-peak distribution of velocities as seen in Fig.~\ref{figure6}f. 
Large clusters exhibit a strong peak at $v_x\sim{0}$, whcih corresponds to particles in clogs. The pdf decreases monotonously after the peak, exhibiting a broad range of intermediate velocities, and a marked depletion of particles moving at $v_x\simeq{1}$. 
Hence, almost all particles are slowed down or trapped in a small number of larger clusters, containing a wide distribution of velocities.

To summarize, the velocity distributions  highlight the nature of the abnormal flow and helps understand how for a given area fraction increasing $\phi_{\text{pin}}$ local clogs arise and affect the flow and system morphology.   
Typically, for $\phi<\phi_{pin}^{a}$ disks flow freely, either in big or small clusters. Above $\phi_{pin}^{a}$ large clusters peak in the distribution around $v_x=0$ due to constrictions and bottlenecks hindering the flow and leading to intermittent flows. Small $\phi$ results in a landscape of uncorrelated clogs and free particles, characterised by a bimodal distribution of velocities. Increasing $\phi$ weakens the bimodal distribution of velocities. Instead, large, dense clusters contribute to increase the correlation of clogging events, resulting in a mixed distribution of velocities that peak at $v_x\sim{0}$

\section{Conclusions}
\label{sec:level6}

We have carried out a thorough study on how forced particles  move and give rise to flow in a randomly disordered obstacle landscape. The methodology put forward has allowed us to identify  and quantify a regime of abnormal flow, where locally clogged regions  persist in time and intermittent motion emerges, from the normal flow regime, where generally, the flow is well defined in the whole disordered system. 
We have classified the  properties of these two regimes at small and large densities, characterized by the development of a   bimodal velocity distribution for small densities, and a large region of coexistence of particles with mixed velocities in large clusters for large densities.  
The weak dependence of the critical anomalous flow regime shows that different area fractions may reach the abnormal regime at different obstacle densities due to cooperation between flowing particles, which fluidize the system and hinder clogged states. The flowing behavior of the forced disks is also altered in the abnormal regime, where the distribution of flow through local regions of the landscape is maximum for arrested clusters and decreases monotonously, in comparison to a non-zero maximum peak in the normal regime. 

We have  characterized some of the structural features related to the anomalous flow regime by analysis the morphologies of particle clusters. We have observed that, independently to the total density, anomalous flows always exhibit a diverging average cluster size, which indicates that there is no characteristic cluster size scale. 
This contrasts with the fact that the abnormal flow density of obstacles,$\phi_{\text{pin}}^a$,  depends weakly on $\phi$, meaning that there is a characteristic obstacle space, $l_a$, which results in clog formation, favouring large densities in bottlenecks and small densities in other regions.
It is for this reason that we observe different dynamical and structural features for small and large local densities: Small densities start with disconnected flows, and large densities starts with connected flows, but both of them separate in large and small density regions in the anomalous regime as  $\phi_{pin}$ increases, translating into an increase of the average cluster size in the first case and a decrease in the second case.

The study  performed has shown that the transition from normal flow to clogging is complex, and it is   controlled by a broad region of abnormal flow where local clogging events coexist with  the underlying flow imposed by the external driving. The nature and magnitude of these events  strongly correlates with the distribution of  particle clusters  that nucleate and develop around local constrictions. 
This correlation is not trivial, since clogs in bottlenecks depend on very specific structural and dynamic properties, such as the bottleneck inclination with respect the force, the number of particles instantaneously arriving to a specific bottleneck and the size of the bottleneck. However, there is still generic features in the abnormal flow, such as the constant $\phi^a_{\text{pin}}$ or separation in large and small density regions.
Therefore, the abnormal regime, initially identified in systems that undergo a clogging transition through a single obstruction, is also present  in a disordered system, characterized  by a spatial distribution of bottlenecks, unifying our understanding of  the transition toward clogging. 

The flexible methodology developed here can be applied to a wide variety of  systems. from heterogeneous mixtures of particles to interacting active matter, to gain insight of how cooperation can be maximised to avoid local clogged states or, inversely, achieve locally spatial flows at some regions of the landscape.

\begin{acknowledgments}
I.P. acknowledges support from Ministerio de Ciencia e Innovaci\'on  MICIN/AEI/FEDER for financial support under grant agreement PID2021-126570NB-100 AEI/FEDER-EU, and from Generalitat de Catalunya  under Program Icrea Acad\`emia and project 2021SGR-673.
\end{acknowledgments}

\nocite{*}

\bibliography{apssamp}

\end{document}